\documentclass[%
 preprint,
 amsmath,amssymb,
]{revtex4-1}

\usepackage{graphicx}
\usepackage{amssymb}
\usepackage{setspace}
\usepackage[utf8]{inputenc}
\usepackage{array}
\usepackage{natbib}
\usepackage{amsmath}
\usepackage{multirow}
\usepackage{hyperref}
\usepackage{subfigure}

\usepackage{color}
\definecolor{redcolor}{rgb}{1.0,0.,0.}

\definecolor{bluecolor}{rgb}{0,0.,1}

\begin{document}

\preprint{}

\title{A comparative analysis of local network similarity measurements: application to author citation networks}

\author{Adilson Vital Jr.$^1$ and Diego R. Amancio$^1$}

\affiliation{
$^1$Institute of Mathematics and Computer Science, University of S\~ao Paulo, S\~ao Carlos, Brazil\\
%
%
%
}

\date{\today}

\begin{abstract}
Understanding the evolution of paper and author citations is of paramount importance for the design of research policies and evaluation criteria that can promote and accelerate scientific discoveries. Recently many studies on the evolution of science have been conducted in the context of the emergent science of science field. While many studies have probed the link problem in citation networks, only a few works have analyzed the temporal nature of link prediction in author citation networks. In this study we compared the performance of 10 well-known local network similarity measurements to predict future links in author citations networks. Differently from traditional link prediction methods, the temporal nature of the predict links is relevant for our approach. Our analysis revealed interesting results. The Jaccard coefficient was found to be among the most relevant measurements. The preferential attachment measurement, conversely, displayed the worst performance. We also found that the extension of local measurements to their weighted version do not significantly improved the performance of predicting citations. Finally, we also found that a neural network approach summarizing the information from all 10 considered similarity measurements was not able to provide the highest prediction performance.

\end{abstract}

\maketitle



\section{Introduction}

Understanding citation patterns is of paramount importance to the understanding of the evolution of science~\cite{Fortunatoeaao0185,nielsen2021global}. Many efforts have been devoted to understand the mechanisms behind citations~\cite{molleri2018towards,liu2021adaptive}. This type of knowledge has allowed an enhanced quantification of evaluation indexes in the \emph{Scientometrics} field~\cite{bai2016identifying}. At the macroscopic level, paper citations are known to be dependent on age, field, journals visibility and other factors~\cite{amancio2012three,krumov2011motifs}. It is well known that citations also are affected by the preferential attachment rule, since more cited papers tend to accrue even more citations. This effect holds for both papers and authors citations~\cite{eom2011characterizing,wang2008measuring,recency}.

Several studies have been devoted to understand the mechanisms underlying citations, but most of them have been limited to analyzing and predicting citation counts~\cite{recency}. In~\cite{amancio2012three} the authors proposed a model that considers three features to predict the behavior of papers citation and authors' h-index. The model considered the preferential attachment rule, the semantical similarity between papers and a memory effect to mimic the tendency of older papers being less cited. While this and other models have been effective to reproduce the \emph{distribution of citations} (and other network measurements), they did not assess the actual correspondence of each \emph{individual edge}. This means that microscopic citation behavior might not be reproduced even though
macroscopic features are consistent with the behavior of real-world citation networks. Similar citation distribution analyses have also been performed at the author level~\cite{recency}.

A more detailed citation analysis considering both end points of a citation can be performed via link prediction techniques~\cite{lu2011link}. A comparison of similarity measurements was performed in the context of predicting links in paper citation networks~\cite{shibata2012link}. The authors found that the Jaccard coefficient and betweenness centrality affect the predictability of the machine learning system. In addition, a dependency on how the fields are organized was reported, since most predictive systems predicts citations within the same field. Temporal link prediction has also been studied in patent citation networks~\cite{patent}. Surprisingly, the study conducted in~\cite{patent} found that structural deep network embedding was not a good measurement for the task of predicting citations between patents.
%

While most works in predicting future citations have been performed at the paper/document level, here we focus on predicting citations between authors. Studying the individual citation behavior of particular interest because it can reveal the emergence of individual citation patterns~\cite{radicchi2009diffusion,Fortunatoeaao0185}.  This type of information can be used not only to evaluation purposes, but can be used to understand how a field evolves~\cite{silva2016using,powell2005network}. Because most citation behavior implies some type of similarity between authors, predicting author citations could also be used to suggest potential effective collaborations~\cite{lande2020link}.

In the context of predicting authors citations, here we carried out a comparison of traditional local network similarity measurements for the task of predicting citations between authors. We conducted our link prediction comparative analysis in a dataset comprising more than $450,000$ papers published in Physics journals. Differently from other studies based on author analysis, our methodology is not impacted by authors' names ambiguity~\cite{zhang2020author,sebo2021accuracy,milojevic2013accuracy,amancio2015topological,nieconstruction}. The considered dataset is enriched with names information extracted from the \emph{Microsoft Academic Graph}, which reports high accuracy in the name disambiguation task.

We limited our comparative analysis to local traditional network measurements for two main reasons: (i) local network measurements can be efficiently computed in very large datasets, with good accuracy results~\cite{10.1371/journal.pone.0181079}. (ii) the same idea of local neighborhood analysis can be extended to include further hierarchies. Thus, quasi-local similarity measurements can be introduced using the same local measurements~\cite{amancio2015topological}.
Our analysis considered local network similarity measurements and their respective definition in weighted networks. Owing to the popularity of machine learning strategies in a myriad of applications, we also evaluated the effectiveness of a neural network model in combining evidence from all the considered similarity measurements.

Several interesting results were observed in our comparative analysis. All local measurements were found to yield a better precision performance when links are evaluated in a longer time window. The number of citations established between authors did not improved the predictability of citations, since the performance observed with unweighted indexes and their respective weighted versions turned out to be similar in several cases. All in all, the best performance was achieved with the Jaccard coefficient. We also found that a combination of all ten similarity network measurements in a neural network approach did not outperform the other approaches. Finally, we also found that the preferential attachment rule should be used in combination with other approaches, since this measurement alone turned out to display a low predictive power in author citation networks.


\section{Methodology} \label{sec:methodology}

This section presents the methodology used in this study. Section \ref{sec:dataset} describes the dataset used to analyze authors citations. Section \ref{sec:net} details the construction of author citation networks. The  measurements used to quantify the similarity between two authors are described in Section \ref{sec:lp}.  A neural network approach to the problem is described in Section \ref{sec:deep}. Finally, we report our comparative analysis in Section \ref{sec:res}. Perspectives for future works are presented in Section \ref{sec:con}.

\subsection{American Physical Society Dataset} \label{sec:dataset}

In our analysis, we used the dataset of papers provided by the \textit{American Physical Society} (APS). The dataset  comprises about $450,000$ articles from several APS journals, including  \textit{Physical Review Letters}, \textit{Physical Review A--E} and \textit {Reviews of Modern Physics}. This dataset has been largely used in several other studies~\cite{bai2020measure,chacon2020comparing,li2019reciprocity,recency}. The dataset comprises citations and additional paper metadata, including DOI, journal name, title, list of authors, affiliations, PACS code and others.
In order to avoid noise from names ambiguity, we used Microsoft Academic Graph (MAG) information to obtain authors' names. This same procedure has been used in~\cite{recency}. Because MAG provides a unique identifier for each author,  we also avoid the name name split issue, i.e. when a single author appear with different names in different publications. In sum, while citations at the paper level are obtained from the APS dataset, we used MAG as an additional dataset to address both name ambiguity and name split issues.

\subsection{Network Construction} \label{sec:net}

Author-citation networks are constructed using the following methodology. Given a time interval, we use information from papers to obtain citation between authors. Two authors $X$ and $Y$ are connected by a citation in a given time interval if a paper co-authored by $X$ cited at least one paper co-authored by $Y$. In the weighted version of the network, edges weight represents how many times $X$ cited $Y$.

\begin{figure}[h]
    \centering
    \includegraphics[scale = 0.75]{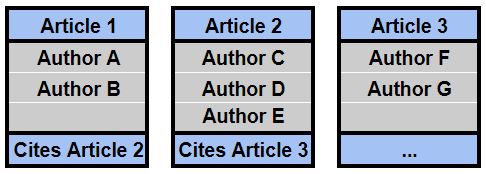}
    \caption{Small set of papers and respective list of authors and citations. The respective author-citation network constructed from this toy dataset is represented in Figure \ref{fig:example}. }
    \label{fig:articles}
\end{figure}

\begin{figure}[h]
    \centering
    \includegraphics[scale = 0.55]{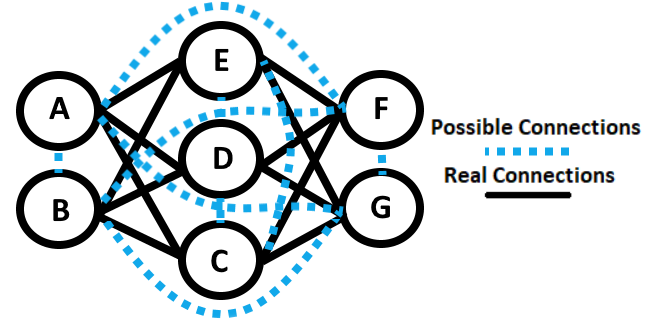}
    \caption{Example of author-citation network constructed from the dataset of papers and respective citations depicted in Figure \ref{fig:articles}. Possible future links that can be predicted in the considered task are represented as blue dashed lines.}
    \label{fig:example}
\end{figure}

Figure \ref{fig:articles} illustrates the process of creating author-citation networks from paper citations. The figure shows e.g. that article 1, co-authored by $A$  and $B$, cites a paper co-authored by $C$, $D$ and $E$. According to this information, the following links between authors are created: $A \rightarrow C$, $A \rightarrow D$, $A \rightarrow E$, $B \rightarrow C$, $B \rightarrow D$ and $B \rightarrow E$. All edges from the toy dataset illustrated in Figure \ref{fig:articles} are depicted in the graph represented in Figure \ref{fig:example} (continuous edges). Note that not all pairs of authors are linked in Figure \ref{fig:example} (see blue dashed lines). These are the potential future links that are evaluated in the link prediction task.

The main advantage of the author citation networks considered here compared is that different from some previous works (see e.g.~\cite{radicchi2009diffusion}) here authors names are disambiguated to avoid noise in the construction of networks.

\subsection{Link prediction} \label{sec:lp}

Once the network is constructed, our aim is to identify future citation between authors. Two frameworks are commonly used for the task~\cite{wang2014link}. The first approach is based on nodes similarity. According to this approach, a similarity value is extracted from all possible links and then sorted in decreasing order. Given a threshold, the considered predicted edges are those taking similarity values above the specified threshold. A different approach consists in considering similarity measurements as features in classification systems. Thus, patterns of links creation are obtained based on previous link creation dynamics. Here we compared the performance of well-known local network similarity measurements. We also used a machine learning method to investigate whether simple local measurements are outperformed by an automatic machine learning strategy.

Our analysis was restricted to the $2\%$ most productive authors observed in the training dataset. The reasons for analyzing only the most productive authors are two-fold: (i) productive authors are the ones most active in the field, so it is expected that they are active along many years. This minimizes the issue of trying to predict links between authors that have stopped publishing after a few papers have been published; and (ii) limiting the study to the most productive authors allows us to analyze the citing behavior of many influential researchers who receive a large fraction of citations in the whole author-citation network~\cite{wang2008measuring}.

In the similarity-based strategy, we computed pairwise similarities between all selected authors. Ten different measurements were used. The similarity computation was performed in the training dataset and then the accuracy of the prediction was evaluated using typical evaluation metrics (see Section \ref{sec:ev} for more details regarding the evaluation). The similarity values were then sorted in decreasing order, and the most similar edges are considered as predicted links according to a threshold value. While we may use the weights in the prediction of links, note that we are not predicting weights but only the existence of a link established in the future.

In the similarity-based approach, the following similarity approaches were used:

\begin{enumerate}

    \item \emph{Common Neighbors} (CN): This is one of the simplest and most used similarity measurements~\cite{newman2001clustering}. It quantifies the total number of shared neighbors. Alternatively, this measurement can be regarded as the number of paths of length 2 connecting two nodes. Mathematically, the similarity $\textrm{CN}(u,v)$ between nodes $u$ and $v$ is computed as
    \begin{equation}
        \textrm{CN}(u,v) = | \Gamma(u) \cap \Gamma(v) |,
    \label{eq:3}
    \end{equation}
    where $\Gamma(v)$ is the set comprising the neighbors of $v$.
    The weighted version of this measurement, defined in \cite{lu2010prediction}, is given by:
    \begin{equation}
        \textrm{CN}^{\textrm{(w)}}(u,v) = \sum_{z \in \Gamma(u) \cap \Gamma(v) } (w_{uz} + w_{vz}), \label{eq:4}
    \end{equation}
    where $w_{uz}$ denotes the weight linking nodes $u$ and $v$.

    \item \emph{Jaccard coefficient} (JC): Another widely used similarity technique is the Jaccard coefficient. This index is similar to CN with the advantage of being normalized in relation to the sum of all neighbors connecting the two data nodes under analysis, i.e.:
    \begin{equation}
    \textrm{JC}(u,v) = \frac{\mid \Gamma(u) \cap  \Gamma(v) \mid}{\mid \Gamma(u) \cup  \Gamma(v) \mid}.
    \label{eq:5}
    \end{equation}
    While in CN two hubs are more likely to share a neighbor just by chance than low-connected nodes. This effect is minimized by the normalization in equation \ref{eq:5}. The weighted version of the Jaccard Index~\cite{desa2011supervised} is given by:
    \begin{equation}
    \textrm{JC}^{\textrm{(w)}}(u,v) = \frac{\sum_{z \in \Gamma(u) \cap \Gamma(v)} (w_{uz} + w_{vz}) }{ \sum_{a \in \Gamma(u)} w_{au} + \sum_{b \in \Gamma(v)} w_{bv} }. \label{eq:6}
    \end{equation}

    \item \emph{Adamic-Adar} (AA): this measurement quantifies the similarity between nodes $u$ and $v$ based on the degree (i.e. the number of neighbors) of nodes in $\Gamma(u) \cap \Gamma(v)$ ~\cite{adamic2003friends}. Mathematically, it is defined as:
    \begin{equation}
        \textrm{AA}(u,v) = \sum_{z \in \Gamma(u) \cap \Gamma(v)}\frac{1}{\log|\Gamma(z)|}.
        \label{eq:7}
    \end{equation}
    Note that nodes in $\Gamma(u) \cap \Gamma(v)$ with higher degrees contribute with a lower weight in the computation of the similarity between $u$ and $v$. The term in the denominator of equation \ref{eq:7} minimizes the contribution of $z \in \Gamma(u) \cap \Gamma(v)$ whenever $z$ is a hub. This is necessary hubs are more likely to be connected to both $u$ and $v$ just by chance. The weighted version of the Adamic-Adar measurement
   ~\cite{lu2010prediction} is defined as
    \begin{equation} \label{eq:8}
    \textrm{AA}^{\textrm{(w)}}(u,v) = \sum_{z \in \Gamma(u) \cap \Gamma(v)} \frac{w_{uz} + w_{vz} }{\log(1 + \sum_{a \in \Gamma(z)} w_{za})}.
    \end{equation}
    %

    \item \emph{Resource Allocation} (RA): Similar to the Adamic-Adar technique, the resource allocation ~\cite{zhou2009predicting} similarity index aims to give lower weight for shared neighbors with a higher degree:
    \begin{equation}
        \textrm{RA}(u,v) = \sum_{z \in \Gamma(u) \cap \Gamma(v) } \frac{1}{| \Gamma(z)|}.
    \label{eq:11}
    \end{equation}
    Notice that here higher degree neighbors contribute with an even lower weight since $\log |\Gamma(z)|$ in equation \ref{eq:7} has now been replaced by $|\Gamma(z)|$ in equation \ref{eq:11}.
    %
    The weighted version of the RA index also punishes neighbors with high strength ($s$)~\cite{lu2010prediction}:
    \begin{equation}
    \textrm{RA}^{\textrm{(w)}}(u,v) = \sum_{z \in \Gamma(u) \cap \Gamma(v) }\frac{w_{uz} + w_{vz}}{s_{z}}.
    \label{eq:12}
    \end{equation}

    \item \emph{Preferential Attachment} (PA): This similarity metric is calculated by the product of the degree of the nodes $u$ and $v$ being analyzed~\cite{Barab_si_2002}:
    \begin{equation}
        \textrm{PA}(u,v) = |\Gamma(u)| \times |\Gamma(v)|.
        \label{eq:9}
    \end{equation}
    Because the preferential attachment states the higher-degree nodes are more likely to accrue new links (see e.g.~\cite{wang2008measuring}), a new link between two highly connected nodes are very likely to appear in the future. The weighted version of the measurement considers the in-strength of nodes instead of the number of neighbors~\cite{desa2011supervised}:
    \begin{equation} \label{eq:10}
        \textrm{PA}^{\textrm{(w)}}(u,v) = \sum_{a \in \Gamma(u)} w_{au} \times \sum_{b \in \Gamma(v)} w_{bv} .
    \end{equation}

    %

\end{enumerate}

%

\subsection{Neural Network Based Approach} \label{sec:deep}

In this method, the prediction is made from a learning model using the similarity measures described in Section \ref{sec:lp} and the instances are the possible new edges that the network may have.
%
In this approach we used the best configuration, i.e. the configuration with highest precision without overfitting~\cite{yegnanarayana2009artificial,nielsen2015neural}.
The best configuration of the neural network was formed by one input layer comprising 10 units and three hidden layers. The first and second layers were formed by 256 and 128 units, respectively. The \emph{relu} function was used in both layers~\cite{yegnanarayana2009artificial,nielsen2015neural}. The last layer comprised one unit.
The input corresponds to the 10 similarity measurements described in Section \ref{sec:lp} and the output is a value ranging between 0 and 1. In other words, the neural network method is a way to summarize all measurements into a single similarity value. Parameter optimization using the procedure described in~\cite{amancio2014systematic}.




%

\subsection{Evaluation} \label{sec:ev}

One of the most traditional means to evaluate the quality of an information retrieval system is to divide the set of edges $E$ into two parts: the training and test edges. The set of training edges will be used to predict the missing edges. A more elaborated method to perform such a division is the $k$-fold cross validation approach~\cite{Kohavi95}. According to this technique, the dataset of links is separated into $k$ different parts (folds) of preferably equal sizes. $k-1$ folds are used for training ($E_{\textrm{train}}$) and the remaining fold ($E_{\textrm{test}}$) is used to test the accuracy of the model. This process must be repeated at $k$ times using different divisions for the test dataset in order to obtain the accuracy of our prediction model according to the average of the accuracy over the repetitions. At each moment, a different division is used as test dataset.

Because author-citation networks represent an evolving dynamic system, an evolution of network structure is expected. Thus, it is natural to expect that new links may appear and old ones may disappear (i.e. no citations between a pair of authors might be observed in the considered period). New nodes can also appear in the network, as new authors are introduced when they publish a paper for the first time.
Given that we are focusing on a link prediction task, we are not predicting links involving new nodes, i.e. nodes that were not observed during the training process.

Owing to the temporal nature of the link prediction problem, we used a modification in the evaluation of the system.
Given an initial year $Y$, we consider a past time window of length $d$ and a future time window of length $p$. Here we aim at predicting links that are formed in our validation database, which consists in all links formed along the interval $t_\textrm{val}$, where $Y < t_\textrm{val} \leq Y+p$. In order to train the algorithms, the training dataset uses the information observed along the interval $t_\textrm{tr}$, where $Y-d \leq t_\textrm{tr} \leq Y$. 

\begin{figure}[h]
    \centering
    \includegraphics[scale = 0.75]{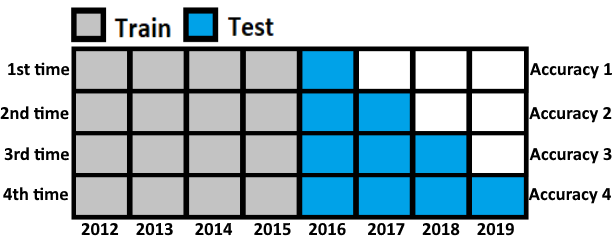}
    \caption{Illustration of the evaluation methodology used in this analysis. In the first evaluation setting, 2012-2015 is used as train dataset and 2016 is used as test dataset. Note that the size of the test dataset increases so that one can evaluate both short- and long-term prediction accuracy.}
    \label{kftemporal}
\end{figure}

In our analysis we varied $p$ so that the prediction quality could be measured at both short- and long- terms. Figure \ref{kftemporal} illustrates the division of the dataset when considering  $Y=2015$ as reference year. In the figure, we also considered $d = 3$ years; therefore the train dataset encompasses the years $2012-2015$. $p$ varies so that the accuracy of the model is evaluated for $p=\{1,2,3,4\}$ years after the reference year $Y$. 

\section{Results and discussion} \label{sec:res}

The comparison of performance is divided into two parts. We first analyze the precision in Section \ref{sec:prec}. The analysis considering the \emph{receiver operating characteristic} (ROC) curve is then discussed in Section \ref{sec:roc}. While the prediction evaluates the accuracy of the model in predicting positive links, the ROC analysis also evaluates the accuracy in not predicting absent future links.

\subsection{Precision analysis} \label{sec:prec}

%
In Figure \ref{fig:precisionsim}, we show the individual behavior of each similarity measurement as more citations are predicted. In our analysis, we considered different test sets. The first one corresponds to the year 2001. This curve is represented as a red curve in Figure \ref{fig:precisionsim}. The largest test set correspond to citations evaluated in the period between {2001} and 2017 (see blue curve). Gray curves correspond to test sets considering papers published in the interval $(2000,y]$, where $2000 < y \leq 2017$.
\begin{figure}[h]
\centering
\includegraphics[scale = 0.8]{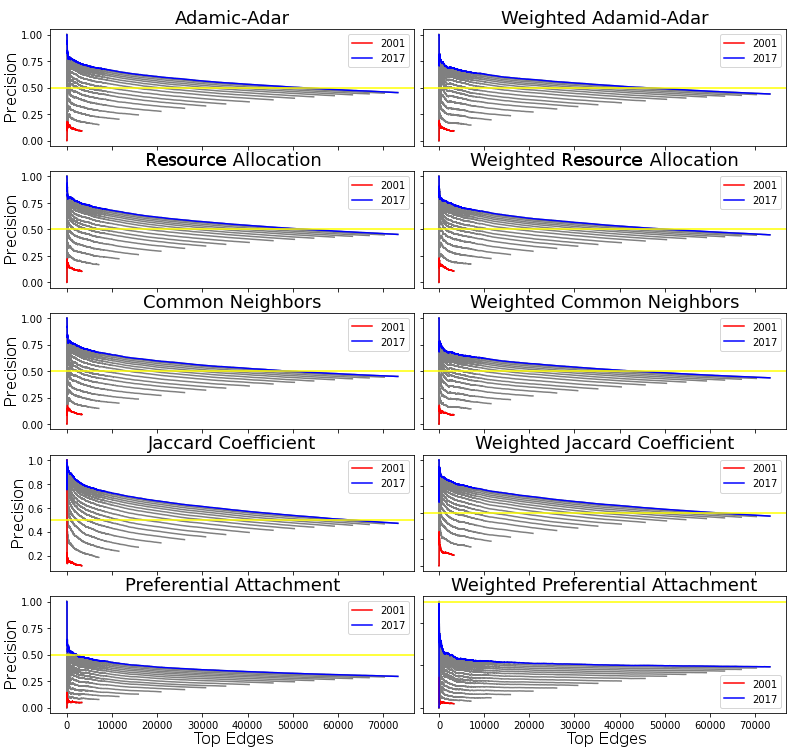}
\caption{Evolution of precision values as additional edges are considered in the analysis. In each subpanel, each curve corresponds to a different test set, or p-fold. The red one corresponds to test set comprising the year 2000 to 2001 alone, while the blue curve corresponds to the test set considering the years $2000-2017$. The yellow line corresponds to the accuracy expected in future links are randomly predicted.}
\label{fig:precisionsim}
\end{figure}

The results in Figure \ref{fig:precisionsim} show that all curves displays similar behavior, meaning that the precision increases as more future edges are evaluated. Therefore, if we consider larger time scale, the tendency is that the most similar authors will indeed be linked by a citation link.
The behavior is also independent of the considered test set and similarity measurements: highly similar edges are predicted with high precision when considering larger periods. The precision slowly drops as similar authors are not linked by citations when considering shorter future time windows.

One interesting finding from Figure \ref{fig:precisionsim} is that both preferential attachment similarity measurements are clearly outperformed by the other measurements. This means that neighborhood information plays an important role in the task of predicting citations in author citation networks. This lack of performance confirms that the prediction of both edge ends in an author citation network is not trivially performed with the preferential attachment rule. We should note, though, that the PA rule is a strong predictor of how many citations a researcher will accrue in the future~\cite{recency}. More specifically, the number of future citations depends mostly on the number of citations received in the last 12-24 months~\cite{recency}. The reasons for not citing similar structural authors are two-fold: authors have a limited vision of the network structure, which may cause them to miss the papers of other authors. While semantical dissimilarity could be a different reason for similar structural authors not citing each other, it should be mentioned that even highly semantical similar papers are frequently overlooked when authors perform a systematic review~\cite{amancio2012using}.

A more detailed comparison of methods, in two distinct test datasets, is shown in Figure \ref{fig:precisionsim}. The precision for selected quantities of included edges (top edges) is also shown in Table \ref{table:final}. The results obtained for the Jaccard Index (both unweighted and weighted versions) turned out to be the most effective measurements to predict future citations. Both methods were significantly better  than the other considered approaches. The normalization provided by the Resource Allocation also yield good results, but with inferior results when compared to the Jaccard coefficient. Surprisingly, the approach based on neural networks achieved only an average performance, being outperformed by the simple Jaccard index in both test datasets. The Adamic-Adar also displayed a low performance. These results reinforces the fact the, when precision is sought, different measurements based on the same information (i.e. common neighbors) can lead to distinct performance. Finally, one can observe that a low precision was observed for the preferential attachment method. Even when predicting the most similar $2,000$ edges, the performance is comparable to a random classification. The weighted version performs even worse. In Table \ref{table:final}, all four precision values for WPA are below 0.25.

\begin{figure}[h]
\centering
\includegraphics[scale = 1]{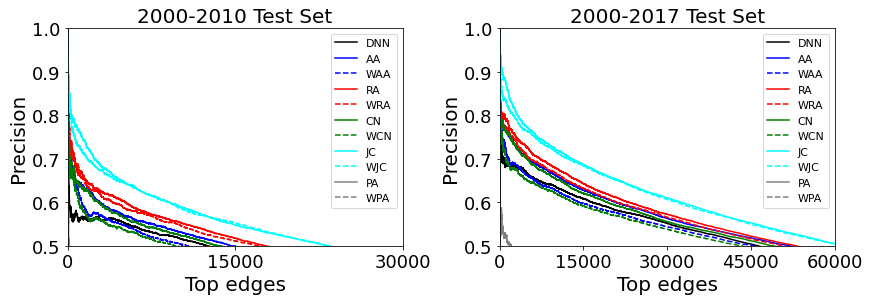}
\caption{Precision comparison when considering future citations in two distinct test datasets. In both test datasets, the highest performance was observed with the Jaccard Index. We also observed only a minor difference between the weighted and weighted version of the considered measurements. Another interesting finding is that the neural network gathering information from all 10 similarity measurements did not yield optimized results. }
\label{fig:precisionsim}
\end{figure}

\begin{table}[htb]
\centering
\caption{Comparison of precision values when considering all the similarity measurements and the neural network model in the 2000-2017 test dataset. Each column represents the precision obtained when different number of edges were included in the link prediction analysis. The total number of included edges varied between $2,000$ to $60,000$ edges.}
\begin{tabular}{c|cccc}
\hline
\hline
\multirow{2}{*}{\bf Method}    & Precision & Precision & Precision & Precision \\
                               &  {\bf $2\times 10^3$}    & {\bf $10\times 10^3$}    & {\bf $20\times 10^3$}   & {\bf $60\times 10^3$}   \\
\hline
{{JC}}              & 0.8315      & 0.7217       & 0.6570       & 0.5054       \\
{{WJC}}             & 0.8055      & 0.7165       & 0.6553       & 0.5042       \\
{{RA}}              & 0.7665      & 0.6796       & 0.6151       & 0.4823       \\
{{WRA}}             & 0.7520      & 0.6674       & 0.6048       & 0.4772       \\
{{AA}}              & 0.7450      & 0.6620       & 0.6042       & 0.4804       \\
{{CN}}              & 0.7445      & 0.6545       & 0.5962       & 0.4754       \\
{{WAA}}             & 0.6890      & 0.6279       & 0.5784       & 0.4659       \\
{{DNN}}             & 0.6850      & 0.6358       & 0.5861       & 0.4654       \\
{{WCN}}             & 0.6755      & 0.6217       & 0.5708       & 0.4615       \\
{{PA}}              & 0.5015      & 0.4252       & 0.3814       & 0.3096       \\
{{WPA}}             & 0.2540      & 0.2193       & 0.2090       & 0.1950       \\ \hline \hline
\end{tabular}
\label{table:final}
\end{table}

\subsection{True vs. False Positive Analysis} \label{sec:roc}

While in the previous section we focused on precision, here we compare the methods by considering both precision and recall in the \emph{receiver operating characteristic} (ROC) curve~\cite{davis2006relationship}. The ROC curve establishes a relationship between true positive and false positive rates. Thus, higher values of AUC are expected whenever true positives are more frequently identified than false positives~\cite{davis2006relationship}
as the threshold in similarity for including new links decreases. This analysis is important because, differently from the precision, the AUC curve also considers the efficiency of the model in \emph{not predicting} links that \emph{will not exist}.

\begin{figure}
\centering
\includegraphics[scale = 0.52]{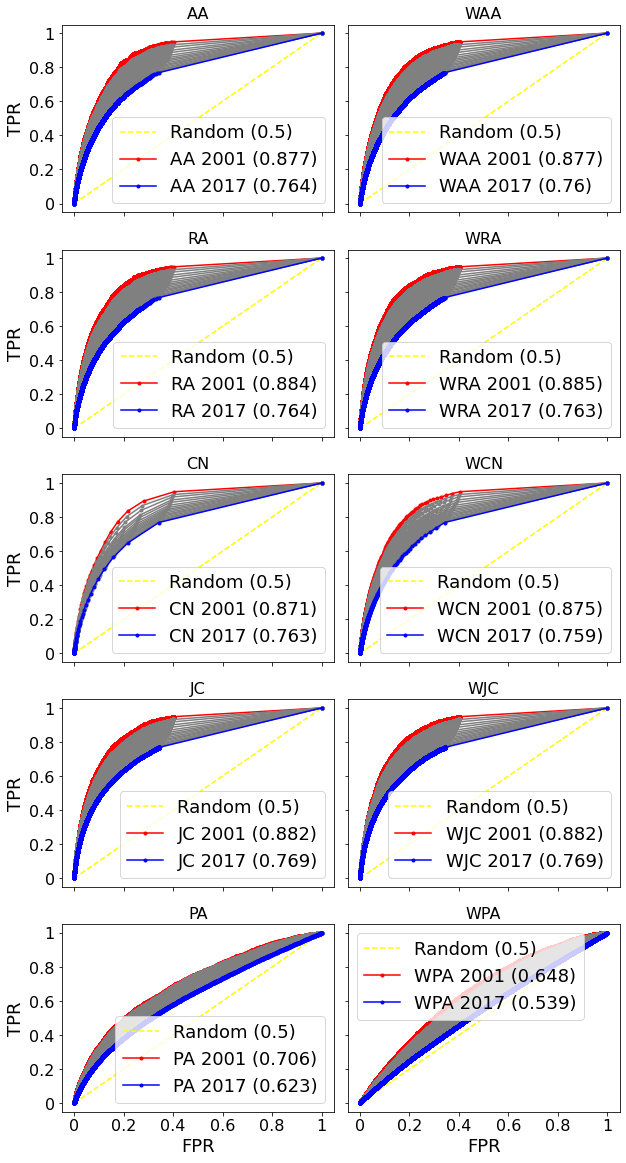}
\caption{\emph{Receiver Operating Characteristic} (ROC) curve for all the similarity measurements considered in our analysis. Overall, the best performance were obtained with the Jaccard index.}
\label{fig:aucsim}
\end{figure}

The results obtained for each of the considered measurements are illustrated in Figure \ref{fig:aucsim}. We show again the results for different test datasets.
We observe that, differently from the precision analysis, a higher performance is observed when predicting links established within short periods. In other words, the efficiency drops when predicting both the existence and absence of links in a long-term scale. %
Interestingly, when considering the largest test dataset (2000-2017), apart from the PA measurement, all measurements have similar performance. The highest differences arise when predicting citations established in the near future. In this context, we do observe a difference in performance. When predicting citations that were established within the first year (2001), the best performances were achieved with RA and JC. This result reinforces the relevance of the Jaccard Index, since both precision and AUC achieved optimized results.
Interestingly this result also reinforces the effectiveness of the Jaccard Index in a wider context, since this same measurement has been reported to be relevant in predicting \emph{paper citations}, even when compared to other global measurements~\cite{shibata2012link}.
Once again preferential attachment measurements turned out to be the ones yielding the lowest performance. Another interesting result is that the weighted versions of the considered measurements were not able to significantly improve the performance of the link prediction task.

When comparing the results in Figure \ref{table:final} with the results obtained with neural networks (result not shown), we found that neural networks achieved only an intermediary performance. In the first considered test dataset (i.e. in 2001), we found a performance of 0.865 for neural networks. This result is outperformed by RA index, which achieved 0.885 in the weighted version. Analogously, for the neural network strategy, the performance observed in the test dataset 2001-2017 was 0.707. This value is clearly outperformed by the Jaccard Index and is equivalent to the PA method. The results confirm once again that the considered neural networks do not outperform more simple indexes based only on the number of shared neighbors.
In other words, the neural network approach is not taking any advantage of the ten local network similarity measurements to summarize different similarity indexes into a single value.
Despite the success of neural networks in particular tasks, our results suggest that there is no advantage of using such measurements as a summarization measurement. This is consistent with recent literature showing that neural networks can also be outperformed by traditional classifiers~\cite{amancio2014systematic}. On the other hand, neural networks and more specifically network embeddings based on neural networks architectures have become popular in recent years. However, they require a broader knowledge of the network structure~\cite{cui2018survey}.

\section{Conclusion} \label{sec:con}

While link prediction have been widely studied in scientometrics scenarios, the analysis and comparisons of methods have been mostly limited to predicting links in citation networks~\cite{daud2017will}. Here we performed a systematic performance comparison of local network similarity measurements in authors citation networks. Because name ambiguity is a major problem when dealing with names in scientometric datasets, we used a disambiguated dataset of names provided by the \emph{Microsoft Academic Graph}~\cite{wang2020microsoft}.

Our comparative analysis focused on local network information to avoid the complexity of analyzing very large networks.
While local measurements indeed may not achieve state of the art performance, given the limited information they rely on, such techniques have shown to yield good performance while not being computationally costly~\cite{shibata2012link}.
In addition to the traditional network similarity measurements, we also used extensions of these measurements that consider the weighted nature of author citation networks. Our comparative approach revealed several interesting results. The Jaccard Index turned out to be the most effective similarity index, when compared with other traditional network similarity measurements. We found that the preferential attachment rule alone is not informative for the task, despite the fact that the total number of citations received by authors is well described by preferential attachment rules~\cite{recency}. Surprisingly, the considered neural network technique did not yield the best results. This suggests that this machine learning strategy, when used to summarize information extracted by local network similarity measurements, is not competitive to predict future citations between authors.
%

This paper focused only on local information to predict new links, since a local analysis mitigates the cost of computing pairwise similarity indexes via global information. In future works, other extensions could be considered in a comparative analysis. One could introduce further hierarchies when comparing neighbors. However, this additional complexity could also lead to noise since many higher-level neighbors might be shared by many authors due to the small-world effect~\cite{hung2010examining}. As a consequence, the characterization performance might decrease with the introduction of deeper concentric circles~\cite{amancio2011using}.  This could be addressed, by providing a lower relevance to higher hierarchies by using a strategy (see e.g.~\cite{amancio2015topological}) that linearly combines similarity values observed in both first and higher hierarchical levels. In a similar direction, information from further hierarchies could be introduced via recent network embeddings techniques, where nodes can be artlessly compared via vector similarity measurements.

An additional comparative analysis could also combine network information with other metada that can not be directly retrieved from author citation networks alone. Previous works have pointed out that some other factors may affect social and information networks in science, including geographical distance, semantical text similarity and other patterns of authors contribution~\cite{amancio2012three,katz1994geographical,wuestman2019geography,stella2020multiplex,stella2019modelling}. All these information could be incorporated into a single model to improve the prediction of the dynamics of citation in author-citation networks.

%
%


\section*{Acknowledgments}

D.R.A. acknowledges financial support from S\~ao Paulo Research Foundation (FAPESP Grant no. 2020/06271-0) and CNPq-Brazil (Grant no. 304026/2018-2). This study was financed in part by the Coordenação de Aperfeiçoamento de Pessoal de Nível Superior -- Brasil (CAPES) -- Finance Code 001.

\bibliographystyle{ieeetr}
\bibliographystyle{abbrv}







\end{document}